\documentclass[%
 reprint,
 amsmath,amssymb,
 aps,onecolumn,nofootinbib,
]{revtex4-2}
\usepackage{multirow}
\usepackage{graphicx}
\usepackage{dcolumn}
\usepackage{bm}
\usepackage{fancyhdr}
\usepackage[colorlinks]{hyperref}
\usepackage[nameinlink,noabbrev]{cleveref}
\hypersetup{
    colorlinks=true,
    linkcolor=blue,
    filecolor=magenta,      
    urlcolor=blue,
   citecolor=blue
}
\usepackage{graphicx}
\usepackage{dcolumn}
\usepackage{bm}
\usepackage{scalerel}
\usepackage{tikz}
\usetikzlibrary{svg.path}

\definecolor{orcidlogocol}{HTML}{A6CE39}
\tikzset{
  orcidlogo/.pic={
    \fill[orcidlogocol] svg{M256,128c0,70.7-57.3,128-128,128C57.3,256,0,198.7,0,128C0,57.3,57.3,0,128,0C198.7,0,256,57.3,256,128z};
    \fill[white] svg{M86.3,186.2H70.9V79.1h15.4v48.4V186.2z}
                 svg{M108.9,79.1h41.6c39.6,0,57,28.3,57,53.6c0,27.5-21.5,53.6-56.8,53.6h-41.8V79.1z M124.3,172.4h24.5c34.9,0,42.9-26.5,42.9-39.7c0-21.5-13.7-39.7-43.7-39.7h-23.7V172.4z}
                 svg{M88.7,56.8c0,5.5-4.5,10.1-10.1,10.1c-5.6,0-10.1-4.6-10.1-10.1c0-5.6,4.5-10.1,10.1-10.1C84.2,46.7,88.7,51.3,88.7,56.8z};
  }
}

\newcommand\orcidicon[1]{\href{https://orcid.org/#1}{\mbox{\scalerel*{
\begin{tikzpicture}[yscale=-1,transform shape]
\pic{orcidlogo};
\end{tikzpicture}
}{|}}}}
\usepackage{amsmath}
\usepackage{amssymb, amsfonts}
\usepackage[greek,english]{babel}
\begin{document}

\preprint{APS/123-QED}

\title{The Other Way Around: From Alternative Gravity to Entropy}


\author{Kamel Ourabah \orcidicon{0000-0003-0515-6728},}\email{kam.ourabah@gmail.com, kourabah@usthb.dz}
\address{Theoretical Physics Laboratory, Faculty of Physics, University of Bab-Ezzouar, USTHB, Boite Postale 32, El Alia, Algiers 16111, Algeria}

\date{\today}

\begin{abstract}
Since the seminal work of Verlinde, the idea that gravity may be an emergent force of entropic origin has gained widespread attention. Many generalizations of this key idea have been considered in the literature, starting from well-known and well-motivated generalized entropies to derive generalized gravity theories. Here, we approach the problem from the opposite direction. We ask whether phenomenologically motivated generalized gravitational theories, yet lacking a strong theoretical justification, may find their origin in an entropic scenario. We examine a set of seven proposals of modified gravity, which have been introduced either (i) as large-scale corrections to Newtonian gravity, aimed at reproducing astrophysical observations in the far field, or (ii) as small-scale corrections, in order to regularize the singularity in the near field. For each proposal, we construct the underlying entropy, producing the desired dynamics in an entropic scenario. This reveals previously unnoticed connections between various proposals. The class of entropies introduced by Sheykhi and Hendi [\href{https://doi.org/10.1103/PhysRevD.84.044023}{Phys. Rev. D \textbf{84}, 044023 (2011)}], exhibiting power-law corrections to the area law, appears to cover a number of useful phenomenological proposals, while the concept of fractional gravity is shown to arise from the recently introduced Barrow entropy. Other entropic forms, involving different type of corrections, also emerge from this procedure. We discuss their implications and their connections with entropies previously introduced in the literature. To broaden our analysis, we extend our discussion to the cosmological context, and examine the effect of these entropies on Friedmann equations.

\end{abstract}

\maketitle


\section{Introduction}
The profound connections existing between gravity and thermodynamics are now common knowledge. The understanding of these connections has quite a long history, dating back to the pioneering works of Bekenstein \cite{1} and
Hawking \cite{2} on black hole physics, and to the work of Jacobson \cite{3}, who showed that the Einstein equation can be regarded as an equation of state. These efforts have progressively opened up a new avenue where gravity can be thought of as an emergent phenomenon \cite{P1,P2,P3,P4} (the very idea of emergence dating back to
Sakharov's “induced gravity” approach \cite{P5,Visser}). Even more profound connections between gravity and thermodynamics have been put forward by Verlinde \cite{Verlinde}. Although similar in spirit to previous works, these connections, hitherto mostly regarded as suggestive analogies, are here taken more literally, identifying therefore a cause, a mechanism, for gravity. The latter is identified with an entropic force, caused by changes in the information associated with the positions of material bodies. By associating the entropy with the Bekenstein-Hawking area law, Newtonian gravity naturally emerges within this picture. To date, while it would be premature to claim that gravity is indeed an emergent phenomenon, one has to recognize that, if true, this picture should have profound consequences. This is the main motivation behind the widespread efforts devoted to understanding this paradigm \cite{VA1,VA2,VA3,VA4}. 

Many generalizations of Verlinde's considerations have been explored in the recent literature. The common idea here is that, instead of the Bekenstein-Hawking area law, one may start from a (well-motivated) generalization thereof, to arrive at a generalization of Newtonian gravity. For example, Sheykhi \cite{Sheykhi2010} considered logarithmic corrections to the area law (motivated by loop quantum gravity \cite{Meissner}), Sheykhi and Hendi \cite{Sheykhi2011} investigated power-law corrections (which may arise if one accounts for excited states \cite{Das2008}), Nicolini \cite{Nicoli} studied the effect of noncommutative geometry corrected
entropies \cite{Nicoli}, Martínez-Merino \textit{et al.} \cite{Obregon} studied the impact of entropies derived from superstatistics\footnote{To be more precise, this is a slightly different formalism from what is known as superstatistics in the statistical mechanics literature \cite{Beck2003}; the parameter usually interpreted as a measure of temperature fluctuations is here identified with the probability itself (see e.g., \cite{Obregon1}).}, and Moradpour \textit{et al.} explored the effect of Rényi and Sharma-Mitall entropies \cite{Moradpour2018}. Alternatively, many authors have also considered a generalization of the equipartition theorem \cite{eqt1,eqt2,eqt3,eqt4} (which, in some formalisms, is related to a generalized form of entropy, through the maximum entropy principle). All these efforts offered a novel way of thinking gravity and its connection with thermodynamics. Here, we propose to take the opposite path; we ask whether well-known phenomenologically motivated generalizations of Newtonian gravity, yet lacking a clear theoretical justification, may find their origin in an entropic scenario.

More precisely, we shall examine a set of seven proposals of generalized Newtonian gravity arguably appearing—given the present state of knowledge—as the most promising alternatives. These can be classified into two different categories, depending on the length scale at which a deviation from Newtonian gravity is expected to take place. In fact, the validity of Newtonian gravity has been probed at length scales ranging from the millimeter \cite{Adel}, in laboratory experiments, to the size of planetary orbits \cite{Fischbach}. Hence, a deviation from Newtonian gravity, if any, should manifest itself either (i) at long ranges (of the order of, or larger than, the size of the solar system), or (ii) at short ranges (typically, below the millimeter)\footnote{This is actually true for tests of general relativity as well, which tightly constrain possible deviations in the solar system,  
but windows remain open for deviations at smaller or larger scales.}. Deviations belonging to the first class propose to modify Newtonian gravity in the far field; they consist of \textit{ad hoc} proposals to deal with the flat rotation curves problem, without invoking dark matter. Deviations of the second type, from another hand,
attempt to eliminate the singularity of the Newtonian force (and potential), at short distances, by weakening the Newtonian interaction in the near field. They generally lack a theoretical justification, since no fundamental description of such a breakdown, yet preserving the
point-particle nature of matter particles, is known to date.

In the first category, we shall examine the gravity model of Maneff \cite{Maneff,Maneff2,Maneff3,Hagihara}, generalizations \textit{à la} Tohline \cite{Tohline} or Kuhn and Kruglyak \cite{Kuhn} (see also \cite{Acedo} for a more modern perspective), the gravitational model of Finzi \cite{Finzi}, and fractional gravity \cite{frac}. Those are among the most studied modifications of Newtonian gravity in the recent astrophysics literature \cite{Astro1,Astro2,Kh,Astro3,Astro4}. At the other extreme, we shall consider small-scale deviations from Newtonian gravity. We examine the case of modified Newtonian gravity in the form of a Yukawa-type potential. This is by now a standard parametrization
of a violation of the Newton inverse-square law, widely used in
experimental tests \cite{Hoyle,Smullin}. We also consider nonlocal gravity of exponential type \cite{Nonlocal1,Nonlocal2,Nonlocal3}, which is a singularity-free extension of Newtonian gravity. In addition, we consider the so-called gradient modification of Newtonian gravity \cite{Lazar}; an alternative procedure to regularize Newtonian gravity at short distances, similar to the
Pauli-Villars regularization \cite{Pauli} in quantum electrodynamics. We deliberately disregard modified Newtonian dynamics (MOND), although it stands among the most promising alternative gravity theories, because its connection with Verlinde's formalism has been widely explored in the literature \cite{eqt2,K2011,K2012,Moradpour}.  

In each case, we construct the underlying entropic form that produces the given gravitational law, in an entropic gravity scenario. This allows unveiling a number of relations between different proposals, hitherto unnoticed—to the best of our knowledge. It appears, for instance, that power-law corrections to the Bekenstein-Hawking entropy, as introduced by Sheykhi and Hendi \cite{Sheykhi2011}, encompass a number of phenomenologically useful theories, and that fractional gravity arises from the so-called Barrow entropy \cite{Barrow}. Other entropic forms, with a more sophisticated correction to the area law, also emerge within this picture. We discuss their potential relevance and their
connection with previously introduced entropies. 

The paper progresses in the following fashion. In Sec. \ref{SecII}, we present a comprehensive overview of the modified gravitational theories that hold our primary focus. In Sec. \ref{SecIII}, we construct the associated entropy for each case, and draw comparisons with previous entropy corrections discussed in the past literature. Progressing to Sec. \ref{SecIV}, our discussion expands to the cosmological context, where we examine the influence of these entropies on the Friedmann equations. Finally, in Sec. \ref{SecV}, we present our concluding remarks and discuss possible avenues for future research.

\section{Generalized gravitational laws}\label{SecII}

For self-completeness, we present in this section the modified gravitational potentials $\Phi$ (produced by a mass $M$) and the associated forces $F$ (felt by a mass $m$), on which we will focus here, and briefly discuss their essential features. These cover generalizations of Newtonian gravity in the far field (Class A) and in the near field (Class B).

\subsection{Large-scale corrections}

The original idea of modifying the $1/R$ gravitational potential can be traced back to Newton himself\footnote{In  \textit{Principia}'s Book I, Article IX, Proposition XLIV, Theorem XIV, Corollary 2, Newton showed that a potential of the form (\ref{1}) leads to a \textit{precessionally elliptic relative orbit}, but he does not study this gravitational law any further. In the 1888-catalogue of the Portsmouth Collection of unpublished manuscripts, however, one can clearly see his efforts to understand the subject.}. In his \textit{Principia}, Newton has already proposed a potential, and the corresponding force, of the form
\begin{equation}\label{1}
\Phi= -\frac{G M}{R} \left(1+\frac{B}{R} \right) \Longrightarrow F= - \frac{G Mm}{R^2} \left(1 - \frac{2B}{R} \right), \quad \text{(Type AI)},
\end{equation}
where $B$ is a constant having the dimension of length. 
Modified gravity theories of this type have been physically justified, later on, by Maneff (sometimes spelled Manev) and others \cite{Maneff,Hagihara}. This model has been used as a possible approximation of general relativity, and succeeded to explain the observed perihelion advance of Mercury. Its main virtue is that it can explain solar-system phenomena, with the same accuracy as relativity, yet using the language of Newtonian mechanics \cite{Diacu}.

Another type of large-scale deviation from Newtonian gravity, widely discussed in the literature, is that of a logarithmically-corrected Newtonian potential, namely

\begin{equation}\label{2}
\Phi=- \frac{GM}{R}-\delta G  Mm \ln (\eta R) \Longrightarrow F = -  \frac{GMm}{R^2} \left ( 1- \delta R \right ), \quad \text{(Type AII)},
\end{equation}
where $\delta$ is a parameter having the dimension of length$^{-1}$ and $\eta$ is an arbitrary constant. This type of generalized gravity has been originally introduced by Tohline \cite{Tohline}, and Kuhn and Kruglyak \cite{Kuhn}. An interesting aspect of potentials in the form of Eq. (\ref{2}) is that they are able to simulate dark matter in astrophysics, and can account for various observational data over different
distance scales. Tohline \cite{Tohline} showed that cold stellar disks can be dynamically stable under potentials of this type, and this phenomenological approach was extended by many authors \cite{Astro4,Kinney,Kirillov}. This is an interesting result, notwithstanding the fact that, so far, there is little
theoretical motivation for such a logarithmic correction\footnote{It has been argued that such a logarithmic correction may originate from quantum effects \cite{Soleng} or from nonlocality \cite{Blome}.}.   Recent comparisons with the data of rotation curves of spiral galaxies indicate that \cite{Acedo} $\delta$ is of
the order of $\delta \sim -0.1 kpc^{-1}$. In our setup, this can be used to constrain the corrections, over the area law, which, in this case, corresponds to the power-law corrections introduced by Sheykhi and Hendi \cite{Sheykhi2011} (see next Section).

Yet another proposal is that of Finzi \cite{Finzi}, which is merely an attempt to solve the problem of the stability of clusters of galaxies. The model proposes to modify the Newtonian potential, so that to produce a stronger attraction at large scales. That is,

\begin{equation}\label{8}
\Phi= - G_*  M \left(\frac{1}{R^{1 / 2}}\right) \Longrightarrow F= - \frac{G_*  Mm}{2 R^{3/2}} \quad(R \gg \mathcal{L}), \quad \text{(Type AIII)},
\end{equation}
where $G_* = 2 k / \mathcal{L}^{1/2}$, $k$ being a constant and $\mathcal{L}$ is a characteristic length $\sim 0.5 kpc$. Contrary to the two cases discussed above, the present model is not, strictly speaking, a correction to Newtonian gravity but a generalization thereof, only valid in the limit of large length scales. Newtonian gravity is nonetheless formally recovered for $R \approx \mathcal{L}$. Admittedly, this proposal has had only little attention, given the success of MOND theory to address the same problem. It is nevertheless interesting to ask how a gravitational theory of this type would emerge in an entropic gravity scenario.

The last type of large-scale deviation from Newtonian gravity we shall consider is that of fractional gravity. Loosely speaking, this kind of theories has quite a long history, and has been investigated by many authors \cite{frac2,frac3,frac4}. Here, we will be mainly interested in the formulation given in \cite{frac}, which has been shown to correctly reproduce the key characteristics of MOND, yet in a linear setup. This approach is based on the fractional Poisson equation, namely
\begin{equation}\label{4}
(-\Delta)^d \Phi=-4 \pi G \ell^{2-2 d} \rho,
\end{equation}
where $\Delta$ denotes the standard Laplacian while $\rho$ and $\ell$ stand for the mass density distribution and a characteristic length scale, respectively. Above, one has $1 \leq d \lesssim 3/2$, and the standard Poisson equation is recovered for $d=1$. By solving Eq. (\ref{4}), one finds the gravitational potential, and the associated force, as follows \cite{frac}

\begin{equation}\label{5}
\Phi=-\frac{\Gamma\left(\frac{3}{2}-d\right)}{4^{d-1} \sqrt{\pi} \Gamma(d)}\left(\frac{\ell}{R}\right)^{2-2 d} \frac{G M}{R} \Longrightarrow F= -\frac{(3-2d) \Gamma\left(\frac{3}{2}-s\right)}{4^{d-1} \sqrt{\pi} \Gamma(d)}\left(\frac{\ell}{R}\right)^{2-2 d} \frac{GMm}{R}, \quad \text{(Type AIV)},
\end{equation}
with\footnote{Eq. (\ref{5}) can be extended to the case $d=3/2$. In this case the potential reads $\Phi = \frac{2}{\pi} \frac{G M}{\ell} \log (R / \ell)$, which should be understood in the regularized sense (see \cite{frac,frac5} for details).} $1 \leq d < 3/2$. Here again, Newtonian gravity is recovered for $d=1$ whereas for $d \to 3/2$, the large-scale behavior predicted by MOND is correctly reproduced \cite{frac}. This places fractional gravity as a promising alternative to the dark matter paradigm, as it naturally encompasses both Newtonian gravity and MOND’s asymptotic behavior.

\subsection{Small-scale corrections}

We move now to deviations from Newtonian gravity in the near field. 
\\

We consider three versions of a small-scale deviation from Newtonian gravity. First, we consider modifications involving a Yukawa-type of potential\footnote{It may be instructive to note that such
potentials naturally occur in non-gravitational media, such as plasmas \cite{Plasma} and Bose-Einstein condensates \cite{Wang}. This may offer new opportunities for the design of laboratory experiments, emulating gravity models of this type, in these media, playing the role of ‘gravity analogs' (see e.g., \cite{Tito, moi1,moi2}).}, with strength $\alpha$ and range $\lambda$, namely

\begin{equation}\label{6}
\Phi= - \frac{GM }{R}(1+ \alpha e^{-  R/ \lambda}) \Longrightarrow F= - \frac{G M m}{R^2} \left[1+ \alpha e^{-  R/ \lambda} \left(1+ \frac{R}{\lambda} \right) \right], \quad \text{(Type BI)}.
\end{equation}
Generalizations of this type have been widely explored in the literature, both at the theoretical and at the experimental levels. Theoretically, they are believed to arise either from coupling to massive particles or from the compactification of extra dimensions \cite{Smullin}. From the experimental side, in a test of
the gravitational inverse-square law below the dark-energy
length scale, Kapner \textit{et al.} \cite{Kapner} found that (for a strength $|\alpha|=1$), one should have $\lambda <56 \mu m$, while a recent experiment by Lee \textit{et al.} \cite{Lee} found $\lambda < 38.6 \mu m$. Note that for $\alpha=-1$, both the potential and the force in Eq. (\ref{6}) do not suffer from the singularity at $R=0$.

Although originally advocated as a small-scale correction to Newtonian gravity, potentials of this form have also been considered as a large-scale deviation from Newtonian, gravity and have been explored in the cosmological context \cite{Almeida,Martino}. This is because, in their weak field
limit, $f(R)$ gravity models show a Yukawa-like correction to the Newtonian
gravitational potential. Solar System tests reveal that \cite{Martino}, for a range $\lambda = 5000 AU$, the
strength must rely in the range $\left[2.70 ; 6.70 \mid \times 10^{-9}\right.$. In an entropic scenario, these bounds provide constrains on the correction over the Bekenstein-Hawking area-law, as discussed next.

Another small-scale deviation from Newtonian gravity is that of nonlocal gravity of an exponential type \cite{Nonlocal1,Nonlocal2,Nonlocal3} (see also discussion in \cite{Lazar}). 
In this case, the gravitational potential and force read as
\begin{equation}\label{A2}
\Phi =-\frac{G M}{R} \operatorname{erf}\left(\frac{R}{2 \ell}\right) \Longrightarrow {F}=-\frac{G M m}{R^2}\left(\operatorname{erf}\left(\frac{R}{2 \ell}\right)-\frac{R}{\sqrt{\pi} \ell} \mathrm{e}^{-R^2 /\left(4 \ell^2\right)}\right),  \quad \text{(Type BII),}
\end{equation}
where
\begin{equation}
\operatorname{erf} (z) :=\frac{2}{\sqrt{\pi}} \int_0^z e^{-t^2} \mathrm{~d} t
\end{equation}
denotes the Gauss error function and $\ell$ is the characteristic length scale parameter of the nonlocal
theory. Potentials of this type reduce to the Newtonian potential for large distances (as compared to the characteristic length scale $\ell$), while the Newtonian interaction is weakened in the near field, allowing for a singularity-free gravitational law.

Another alternative, for the regularization of gravity at short distances, is that of (second order) modification of Newtonian gravity \cite{Lazar}, which is a good approximation of nonlocal gravity of the exponential type (i.e., Type BII). In this framework,
the gravitational potential field remains local but it satisfies a
partial differential equation of sixth order, including internal
characteristic length scales $a_1$ and $a_2$, determining the range of modification
of Newtonian gravity. 
In this case, the potential and the corresponding force read
\begin{equation}\label{A4}
\Phi=-\frac{G M}{R} f_0\left(R, a_1, a_2\right) \Longrightarrow F=-\frac{G Mm}{R^2} f_1 \left(R, a_1, a_2\right), \quad \text{(Type BIII)}, \end{equation}
where, in the real case (i.e., for real and
distinct length scale parameters $a_1$ and $a_2$), the auxiliary functions $f_{0,1}$ read as \cite{Lazar}

\begin{equation}
\begin{aligned}
f_0\left(R, a_1, a_2\right)=&1-\frac{1}{a_1^2-a_2^2}\left[a_1^2 \mathrm{e}^{-R / a_1}-a_2^2 \mathrm{e}^{-R / a_2}\right], \\
f_1\left(R, a_1, a_2\right)= & 1-\frac{1}{a_1^2-a_2^2}\left[a_1^2 \mathrm{e}^{-R / a_1}-a_2^2 \mathrm{e}^{-R / a_2}\right] 
 -\frac{R}{a_1^2-a_2^2}\left[a_1 \mathrm{e}^{-R / a_1}-a_2 \mathrm{e}^{-R / a_2}\right]. 
\end{aligned}
\end{equation}
The
gravitational potential in this case read as a superposition
of a long-range Newtonian potential and a short-range bi-Yukawa-type potential\footnote{Here again, this type of bi-Yukawa potentials naturally arise in non-gravitational media, such as plasmas \cite{Plasma2005}.}. The gravitational potential and the associated force [viz. Eq. (\ref{A4})] are singularity-free and show a modification at short distance, while they converge to their Newtonian counterparts in the far field.

In these last two cases (Type BII and BIII), to the best of our knowledge, no experimental bounds are known to date but possible tests, involving gravity analogs, may be accessible in state-of-the-art laboratory experiments (see \cite{moi1}).

\section{From generalized gravity to generalized entropy}\label{SecIII}

Now that we have provided an overview of the gravity models of interest, our focus shifts to constructing the underlying entropic form for each case. In doing so, we will follow the standard procedure (see e.g., \cite{Sheykhi2010,Sheykhi2011,Nicoli} for more details and intermediate steps).

According to Verlinde’s argument, when a test particle moves apart from the holographic screen—a storage device of information, the magnitude of the entropic force on this body satisfies the relation
\begin{equation}
F \triangle x=T \triangle S,
\end{equation}
where $\Delta x$ denotes the displacement of the particle from the holographic screen while $T$ and $\Delta S$ stand respectively for the temperature and the entropy change on the screen. Upon identifying the entropy with the area law, namely\footnote{Throughout this section, we set the Boltzmann constant $k_B=1$.}
\begin{equation}\label{arr}
S= \frac{A}{4 \ell_p^2},
\end{equation}
where $A=4 \pi R^2$ is the area of the horizon and $\ell_p=\sqrt{G \hbar / c^3}$ is the Planck length, Newtonian gravity emerges \cite{Verlinde}. Instead of the area law, we consider here a generic form of entropy $S(A)$, i.e., an arbitrary function of $A$.
This function may incorporate universal parameters, such as characteristic length scale parameters, and in a specific limit of these parameters, it converges to the standard area law (\ref{arr}).

The rest of the argument is quite standard.
We consider a test mass $m$ in the gravitational field produced by a source mass $M$, located at the center of a spherically symmetric surface $\mathcal{S}$. The test mass $m$ is assumed to be very close to the surface, as compared to its reduced Compton wave-length $\lambda_m \equiv \hbar / mc$. When the test mass $m$ is a distance 
$\Delta x = \eta \lambda_m$ away from the surface $\mathcal{S}$, the entropy of the surface changes by one fundamental unit 
$\Delta S$ fixed by the discrete spectrum of the area of the surface via 
\begin{equation}
\Delta S=\frac{\partial S}{\partial A} \triangle A.
\end{equation}
Now, we (i) identify the energy of the surface with the mass $M$ (i.e., $E=M c^2$), (ii) assume that the number of “bits” of information living on the surface $\mathcal{S}$ is proportional to the area $A$, and (iii) assume the equipartition theorem. It is then a simple exercise to find the resulting force, which reads as (see e.g., \cite{Sheykhi2010,Sheykhi2011})
\begin{equation}\label{F}
F=-\frac{4 \ell_p^2 G M m}{R^2} \left. \frac{\partial S}{\partial A} \right |_{A=4 \pi R^2} ,
\end{equation}

which can be inverted to return the entropy $S$, given the force $F(R)$, namely
\begin{equation}\label{S}
S=- \frac{1}{16 \pi \ell_p^2 G M m} \int \left. F(R) \right|_{R=\sqrt{A/4 \pi}} A dA.
\end{equation}
A word of caution is in order here. Note that such a simple inversion is possible here because we are considering corrections in the force involving only $R$, but conserving the same dependency on the masses (i.e., $F$ remains proportional to the product of masses $Mm$). Indeed, this holds true for the set of proposals we are examining here (see Sec. \ref{SecII}). In this case, the entropy (\ref{S}) is mass-independent, as required. It is only a function of $A$, and some length scale parameters that are regarded as fundamental constants. If the force is no longer proportional to $Mm$ (as e.g. the case discussed in Ref. \cite{eqt1}), constructing the corresponding entropy becomes a nontrivial matter involving a modification of the equipartition theorem.

Using Eq. (\ref{S}), we are in position to construct the entropic form associated with each of the gravitational models discussed in Sec. \ref{SecII}. We start with forces exhibiting a large-scale deviation from Newtonian gravity (Class A). As will be shown, their associated entropies correspond to simple and well-known corrections. Short-range deviations from Newtonian gravity (Class B) exhibit more complicated corrections, and will be discussed next.

\subsection{Large-scale corrections}

For a gravitational force \textit{à la} Maneff (Type AI) [viz. Eq. (\ref{1})], using Eq. (\ref{S}), one easily finds the associated entropy $S$ which reads as  
\begin{equation}\label{15}
S= \frac{ 1}{4 \ell_p^2}\left[A-8 B \sqrt{\pi} A^{1/2}\right], \quad \text{(Type AI)}.
\end{equation}
Similarly, for a gravitational force of Type AII  [viz. Eq. (\ref{2})], one finds 
\begin{equation}\label{16}
S= \frac{ 1}{4 \ell_p^2}\left[A+ \frac{2 \delta}{3 \sqrt{4 \pi} }A^{3/2} \right], \quad \text{(Type AII)}.
\end{equation}
It is instructive to note that both entropic forms (\ref{15}) and (\ref{16}) are special cases of the entropic law, introduced by Sheykhi and Hendi \cite{Sheykhi2011}, characterized by a power-law correction over the area law, that is

\begin{equation}\label{17}
S=\frac{ 1}{4 \ell_p^2}\left[A-K_\alpha A^{2-\alpha / 2}\right],
\end{equation}
where
\begin{equation}
K_\alpha=\frac{\alpha(4 \pi)^{\alpha / 2-1}}{(4-\alpha) r_c^{2-\alpha}},
\end{equation}
$r_c$ being a characteristic length scale. Both the gravitational model of Maneff \cite{Maneff,Hagihara} (Type AI) and that of Tohline, Kuhn and Kruglyak \cite{Tohline,Kuhn} (Type AII) can be regarded as special cases, emerging from power-law corrections over the area law, in an entropic gravity scenario. The entropy associated with the gravitational force of Type AI is recovered for $\alpha \equiv 3$ and $B \equiv 3 r_c/4$, while that associated with a force of Type AII is obtained for $\alpha \equiv 1$ and $\delta \equiv - 1/2r_c$. In this last case, one may take advantage of the recent comparison with the data of rotation curves of spiral galaxies \cite{Acedo} pointing toward the value $\delta \sim -0.1 kpc^{-1}$, to find an estimate for $r_c$. This leads to $r_c \sim 5 kpc$. For this value of $r_c$ (and $\alpha=1$), the entropic form of Sheykhi and Hendi (\ref{17}) correctly reproduces the empirical success in the description of the rotation curves of spiral galaxies, in an entropic scenario.

Proceeding similarly, one finds the entropic form corresponding to a gravitational force \textit{à la} Finzi (Type AIII) [viz. Eq. (\ref{8})] which reads as

\begin{equation}\label{19}
S= \frac{ 1}{4 \ell_p^2}\left[ \frac{4k}{5 G (4 \pi)^{1/4} \sqrt{\mathcal{L}}} A^{5/4} \right], \quad \text{(Type AIII)},
\end{equation}
and, more generally, for fractional gravity (Type BIV) [viz. Eq. (\ref{5})], which reads
\begin{equation}
S= \frac{ 1}{4 \ell_p^2}\left[ \frac{2^{5-4d} \pi^{1/2-d} \Gamma (5/2-d) }{\Gamma (1+d) \ell^{2d-2}} A^d \right], \quad \text{(Type AIV)}.
\end{equation}
It is instructive to note that, in this case, the corresponding entropies coincide with the so-called fractal entropy introduced by Barrow, namely \cite{Barrow}
\begin{equation}\label{Barrow}
S=\left(\frac{A}{4 \ell_p^2}\right)^{1+\Delta / 2},
\end{equation}
where $0 \leq \Delta \leq 1$ quantifies the departure from the Bekenstein-Hawking area law. In particular, $ \Delta = 0$ (equivalent to $d=1$) reproduces the standard area law, while $\Delta = 1$ (equivalent to $d=3/2$) corresponds to the maximal deformation. Note that the Barrow entropy (\ref{Barrow}) is a modification of the area law, arising from the fractal structure of the horizon. The underlying idea behind Eq. (\ref{Barrow}) involves considering a core sphere of the horizon with smaller spheres attached to it, forming a fractal structure. This fractal consists of progressively smaller spheres, similar to well-known fractal patterns like the \textit{Koch snowflake} or the \textit{Sierpi\'{n}ski gasket}. By summing up the surfaces of this hierarchical system, one arrives at an effective area
\begin{equation}
A_{\operatorname{eff}}= 4 \pi R_{\operatorname{eff}}^2,
\end{equation}
with an effective radius defined as
\begin{equation}
R_{\operatorname{eff}} \equiv R^{1+\Delta / 2}.
\end{equation}
The resulting effective surface area is bounded from below by the area of a sphere $A$ (i.e., $\Delta=0$) and from above by $\Delta=1$, in which case the area acts as a volume. The limit of maximal deformation ($\Delta=1$) corresponds \textit{precisely} to the limit $d=3/2$ imposed in fractional gravity, by the very mathematical definition of the fractional Laplacian operator (see e.g., \cite{Lap1,Lap2}). This suggest a deep connection between the concept of fractional gravity and the fractal Barrow entropy.

An alternative interpretation, formally equivalent to Barrow fractal entropy, is that of the so-called Tsallis (or better Tsallis-Cirto) entropy \cite{Tsallis1,Tsallis2}, i.e., $S_{\delta} \sim A^{\delta}$, which, although formally equivalent to Eq. (\ref{Barrow}), has a very different interpretation\footnote{The rationale here is that, while the Bekenstein-Hawking entropy is not extensive for a $3d$ object (i.e., it is proportional to the area $A$), a generalized entropic form $S_{\delta} \sim A^{\delta}$ can be made extensive (proportional to the volume), for $\delta=3/2$. More generally, for a $d$-dimensional system ($d>1$), the extensive nature of entropy is preserved if one defines $\delta = d/(d-1)$.}. Recent studies on modified Friedmann cosmology, based on Tsallis-Cirto entropy, provide the upper bound \cite{up} $\delta \leq 2$, which is compatible with Type AIII gravity ($\delta \equiv 5/4$) and, more generally, with Type AIV gravity ($\delta \equiv d$, with $1 \leq d \leq 3/2$).

\subsection{Small-scale corrections}

We move now to small-scale deviations from Newtonian gravity (Class B). For a deviation in the form of a Yukawa potential (Type BI) [viz. Eq. (\ref{6})], one obtains using Eq. (\ref{S}), the following entropic form

\begin{equation}\label{A25}
S= \frac{ 1}{4 \ell_p^2}\left[  A- 2 \alpha \lambda^2 e^{- \sqrt{\frac{A}{4 \pi \lambda^2}}}\left(12 \pi + 6 \sqrt{\frac{\pi A}{\lambda^2}}+ \frac{A}{\lambda^2} \right)\right], \quad \text{(Type BI)},
\end{equation}
which reduces to the standard area law for $\alpha=0$, as required. Moreover, for large $A$, the entropic correction vanishes, due to the decreasing exponential function. Understood as a small-scale effect, one should have a length scale $\lambda$ below the micrometer, hence $\lambda \ll \sqrt{A}$, resulting in a small correction. At the other extreme, in cosmological applications, the range $\lambda$ is much larger but the correction is attenuated by a very small value of the strength $\alpha$ (typically $\alpha \sim 10^{-9}$ \cite{Martino}). To the best of our knowledge, this type of correction to the area law has not been considered in the past literature. We note however that extensions of the area law involving exponential corrections have been considered in Ref. \cite{Obregon}, based on the concept of superstatistics. Comparable exponential corrections have also been shown to arise if microstate counting is carried out for quantum states residing only on the horizon \cite{expp}. Moreover, some interesting computations in string theory show similar exponential corrections as well \cite{string}.

For nonlocal gravity of exponential type (Type BII) [viz. Eq. (\ref{A2})], one finds the following entropy
\begin{equation}\label{26}
S= \frac{ 1}{4 \ell_p^2}\left[  12 \ell  \sqrt{A} e^{-\frac{A}{16 \pi \ell^2}} + \left(A- 24 \pi \ell^2   \right) \operatorname{erf} \left (\sqrt{\frac{A}{16 \pi \ell^2}} \right) \right], \quad \text{(Type BII)},
\end{equation}
which can be regarded as a nonlocal extension of the area law. It can be easily seen that the standard area law $S= A/4 \ell_p^2$ is correctly recovered from Eq. (\ref{26}) for large $A$ (as compared to $\ell^2$). In fact, in this limit, the first term into brackets vanishes, due to the decreasing exponential function, whereas the second term reduces to $A$ (note that $\lim_{x\to \infty} \operatorname{erf}(x) =1$). As far as we know, this type of extension to the area law, involving the error function $\operatorname{erf (z)}$, has not been considered in the literature.

Finally, let us consider the case of gradient field gravity (Type BIII) [viz. Eq. (\ref{A4})]. In this case, the entropy is found as

\begin{equation}
S= \frac{ 1}{4 \ell_p^2}\left[ A+ \frac{1}{a_1^2-a_2^2} \left \{ a_1^2 \left[4 a_1^2 \left (\sqrt{\frac{\pi A}{a_1^2}} + 2 \pi \right) e^{-\sqrt{{A}/{4 \pi a_1^2}}} \right]  -  a_2^2 \left[4 a_2^2 \left(\sqrt{\frac{\pi A}{a_2^2}} + 2 \pi \right) e^{-\sqrt{{A}/{4 \pi a_2^2}}} \right] \right \} \right], \quad \text{(Type BIII)},
\end{equation}
which is similar to entropic corrections in the form of (\ref{A25}), but involving two length scales. For $a_1=a_2=0$, the standard area law is recovered. Moreover, for large $A$, the correction terms vanish, due to the decreasing exponential functions, and the standard area law is correctly reproduced.  

\section{First Law and MODIFIED FRIEDMANN EQUATIONS}\label{SecIV}

In this section, we extend the previous discussion to the cosmological context, and derive modified Friedmann equations, associated with the entropic forms presented in Sec. \ref{SecIII}. In fact, once regarded as a consequence of a general form of entropy, the implications of these alternative gravitational theories on cosmology can be straightforwardly addressed, by using their associated entropy and the first law of thermodynamics. We follow in this derivation the standard lines of reasoning (see e.g., \cite{Sheykhi2011,Sheykhi2010bis,Karami2011}), applied to the entropic forms obtained in Sec. \ref{SecIII}. Throughout this section, we set $\hbar = c= k_B=1$, while we keep the gravitational constant $G$ explicit, such that $\ell_p^2 \equiv G$.

We assume the background spacetime to be spatially homogeneous and isotropic, which is described by the following line element

\begin{equation}
d s^2=h_{\mu \nu} d x^\mu d x^\nu+R^2\left(d \theta^2+\sin ^2 \theta d \phi^2\right),
\end{equation}
with $R=a(t) r$ ($a$ being the scale factor), $x^0=t$, $x^1=r$, and the two-dimensional
metric $h_{\mu \nu}=\operatorname{diag}\left(-1, a^2 /\left(1-k r^2\right)\right)$, were $k$ denotes
the curvature of space. Specifically, $k=0,1,-1$ correspond
to \textit{open}, \textit{flat}, and \textit{closed} universes, respectively. A simple calculation gives the apparent
horizon radius for the Friedmann-Robertson-Walker
(FRW) universe as \cite{FRW1,FRW2}
\begin{equation}\label{R}
R=a r=\frac{1}{\sqrt{H^2+k / a^2}},
\end{equation}
where $H= \dot{a}/a$ is the Hubble parameter. We assume
the matter source in the FRW universe to be a perfect fluid with stress-energy tensor 
\begin{equation}
T_{\mu \nu}=(\rho+p) u_\mu u_\nu+p g_{\mu \nu},
\end{equation}
where $\rho$ and $p$ stand for the mass density and the pressure, respectively. This leads to the continuity equation
\begin{equation}\label{conti}
\dot{\rho}+3 H(\rho+p)=0.
\end{equation}
From another hand, the temperature associated with the
apparent horizon can be defined as \cite{T1,T2}
\begin{equation}\label{xx1}
T=- \frac{1}{2 \pi R}\left(1-\frac{\dot{R}}{2 H R}\right),
\end{equation}
where one has to impose ${\dot{R}} 
\leq {2 H R}$ to ensure that the temperature is positive\footnote{Despite numerous theories and experiments proposing the existence of negative absolute temperature, stringent arguments \cite{tem} suggest that such claims are unfounded, and arise from the use of an inconsistent entropy definition.}. Now, one can derive (generalized) Friedmann equations by applying the first law of thermodynamics\footnote{A slightly different alternative is to write the first law as $T d S=-d E$, with $-dE$ being the heat flux crossing the apparent horizon within an infinitesimal interval of time $dt$ (see e.g. Ref. \cite{Karami2011}).}
\begin{equation}\label{xx2}
d E=T d S+W d V,
\end{equation}
where $W$ stands for the work, and reads as \cite{Hay}
\begin{equation}\label{xx3}
W=\frac{1}{2}(\rho-p).
\end{equation}
Assuming that $E= \rho V$ is the total energy content of the universe inside a $3$-sphere of radius $R$, where $V=4 \pi R^3/3$ is the volume enveloped by the $3$-dimensional sphere, with an area of the apparent horizon $A=4 \pi R^2$, we write
\begin{equation}\label{xx4}
d E=4 \pi \rho R^2 d R = 4 \pi H R^3(\rho+p) d t,
\end{equation}
where we have used the continuity equation (\ref{conti}). Now, combining Eqs. (\ref{xx1})-(\ref{xx4}), and keeping the definition of the entropy $S(A=4 \pi R^2)$ general, we have 
\begin{equation}
H(\rho+p) d t = \left (\left. \frac{\partial S}{\partial A} \right|_{A=4 \pi R^2} \right ) \frac{dR}{\pi R^3}.
\end{equation}
Using the continuity equation (\ref{conti}), and integrating, we get
\begin{equation}\label{xx37}
\rho = - \frac{3}{\pi} \int \left(\left. \frac{\partial S}{\partial A} \right|_{A=4 \pi R^2} \right ) \frac{dR}{R^3} +C,
\end{equation}
where $C$ is a constant of integration, to be determined later by imposing that the standard Friedmann equation is recovered in the proper limit. Eq. (\ref{xx37}) is very general, and valid for an arbitrary entropic form $S(A)$. Now, we focus on the entropic forms obtained in Sec. \ref{SecIII} and use Eq. (\ref{xx37}) to derive the corresponding Friedmann equations. We start with the entropies associated with deviations from Newtonian gravity in the far field (Class A), as they hold more interest in the cosmological context. We study the entropic corrections associated with Class B theories, next.  

\subsection{Large-scale corrections}

Let us start with the class of entropies associated with power-law corrections over the area law (\ref{17}). Those encompass the gravitational law of Type AI [viz. Eq. (\ref{1})], and that of Type AII  [viz. Eq. (\ref{2})] (see the corresponding entropies, Eqs. (\ref{15}) and (\ref{16})). Using Eqs. (\ref{17}) and (\ref{xx37}), one obtains  
\begin{equation}
H^2+\frac{k}{a^2}-r_c^{-2}\left[r_c^a\left(H^2+\frac{k}{a^2}\right)^{a / 2}-1\right]=\frac{8 \pi G}{3} \rho.
\end{equation} 
This corresponds to the generalized Friedmann equation, involving power-law entropic corrections, derived in Ref. \cite{Sheykhi2011,Karami2011}.
The special case of Type AI gravitational law (\ref{15}) corresponds to $\alpha \equiv 3$ and $B \equiv 3 r_c/4$, while that of AII law (\ref{16}) corresponds to $\alpha \equiv 1$ and $\delta \equiv - 1/2r_c$.

From another hand, for the gravitational law of Type AIII [viz. Eq. (\ref{8})] and, more generally,  for fractional gravity (Type AIV) [viz. Eq. (\ref{5})], both formally corresponding to Barrow entropy (\ref{Barrow}), one obtains the following corrected Friedmann equation
\begin{equation}\label{FB}
\left (  \frac{2+ \Delta}{2- \Delta} \right )  \left ( H^2 + \frac{k}{a^2} \right )^{1- \Delta /2} =   \frac{2 (4 G)^{1+\Delta/2} \pi \rho}{3},
\end{equation}
which has been previously derived in Ref. \cite{Sheykhi2021}. In particular, for $\Delta =1/2$, Eq. (\ref{FB}) corresponds to the gravitational law of Type AIII (Finzi gravity), while upon identifying $\Delta \equiv 2 d- 2$, it corresponds to fractional gravity (Type AIV), and constitutes a natural extension thereof to the cosmological context.

\subsection{Small-scale corrections}

Now, let us turn our attention to the entropic forms associated with Class B theories. We begin with the gravitational law of a Yukawa-type (Type BI) [viz. Eq. (\ref{6})]. This is the most interesting law of Class B to address in the cosmological context, since it has not only been advocated as a small scale modification of Newtonian gravity but also as a large scale deviation (see e.g. \cite{Almeida,Martino}). Using the corresponding entropic form (\ref{A25}) and Eq. (\ref{xx37}), one arrives at the following generalized Friedmann equation
\begin{equation}\label{FY}
\left(H^2 + \frac{k}{a^2}  \right) - \frac{\alpha}{\lambda^2} \left [ \lambda \left(H^2 + \frac{k}{a^2} \right) \operatorname{e}^{\frac{1}{\lambda \left(\sqrt{H^2 + {k}/{a^2}} \right) }} \left(- \lambda + \frac{(1-2 \sqrt{\pi})}{\sqrt{H^2 + k/ a^2}} \right) + (1-2 \sqrt{\pi}) \operatorname{Ei} \left(- \frac{1}{\lambda \sqrt{H^2 + k / a^2}} \right) \right] = \frac{8 \pi G \rho}{3},
\end{equation}
where 
\begin{equation}
\operatorname{Ei}(x) :=-\int_{-x}^{\infty} \frac{\mathrm{e}^{-t}}{t} \mathrm{~d} t=\int_{-\infty}^x \frac{\mathrm{e}^t}{t} \mathrm{~d} t 
\end{equation}
is the exponential integral special function. To the best of our knowledge, similar corrections to the Friedmann equation have not been previously reported. Eq. (\ref{FY}) may provide new perspectives on cosmological models involving Yukawa-type gravity (see e.g., \cite{yuk1,yuk2}). One may easily check that for $\alpha=0$, Eq. (\ref{FY}) reduces to the standard Friedmann equation, as required.

Following the same procedure, one may derive entropy-corrected Friedmann equations, associated with nonlocal gravity (Type BII) [viz. Eq. (\ref{A2})] and gradient gravity (Type BIII) [viz. Eq. (\ref{A4})]. Although the corrections in this case are expected to be negligible at the cosmological scale, it remains interesting to formally derive the corresponding Friedmann equations. For Type BII gravity, one obtains 

\begin{equation}\label{AZE42}
\begin{aligned}
&\frac{(H^2+ k/ a^2)}{2 \ell} \left [ \frac{1}{\sqrt{H^2+ k/ a^2}} \left ( \frac{1}{\ell \sqrt{H^2+ k/ a^2}} \frac{2 \operatorname{e}^{- 1 / 4 \ell^2 (H^2+ k/ a^2)}}{\sqrt{\pi}} \right) \right . \\
& \left. + \left( \frac{1}{\ell (H^2+ k/ a^2)} - 2 \ell   \right) \operatorname{erf} \left( \frac{1}{2 \ell \sqrt{H^2+ k/ a^2}} \right) \right] = \frac{8 \pi G \rho}{3},
\end{aligned}
\end{equation}
while for BIII graviy, one has

\begin{equation}\label{AZE43}
(H^2 + k/a^2) + \frac{4 (\Tilde{f_1} - \Tilde{f_2})}{a_1^2-a_2^2}    = \frac{8 \pi G \rho}{3},
\end{equation}
where we have defined the following auxiliary functions
\begin{equation}
\Tilde{f_i} := \left({a_i}{(\sqrt{H^2+k/a^2}}) - {a_i^2}{(H^2+k/a^2)} \right) \operatorname{e}^{- \frac{1}{a_i (\sqrt{H^2+k/a^2})}}+ \operatorname{Ei} \left( - \frac{1}{a_i (\sqrt{H^2+k/a^2})} \right), \quad (i=1,2).
\end{equation}
Here again, one may easily check that the standard Friedmann equation is correctly recovered for a vanishing length scale, i.e. $\ell \to 0$ (for Eq. (\ref{AZE42})), and $a_1 \to0 \wedge a_2 \to 0$ (for Eq. (\ref{AZE43})). 

\section{Conclusion}\label{SecV}

In this paper, we attempted to offer a fresh perspective on the framework of entropic gravity. Unlike previous literature \cite{Sheykhi2010,Meissner,Sheykhi2011,Nicoli,Obregon}, which focused on deriving alternative gravitational laws from a generalized entropic form, we have explored whether alternative gravitational theories, motivated by phenomenological considerations but lacking a strong theoretical basis, can be explained within an entropic scenario. We have examined a set of seven proposals, which have been introduced either (i) as large-scale corrections to the gravitational field, in order to explain astrophysical observations without invoking dark matter, or (ii) as small-scale corrections, to address the singularity problem in the near field. For each case, we have constructed the underlying entropic form, leading to the given gravitational law in an entropic gravity scenario.

This alternative path not only produces (potentially relevant) new entropic forms, but also reveals deep connections between different proposals. Notably, the gravitational models proposed by Maneff \cite{Maneff,Hagihara} and Tohline, Kuhn, and Kruglyak \cite{Tohline,Kuhn}, widely used in fitting astrophysical observations, are shown to emerge as special cases of the power-law corrected entropies introduced in Ref. \cite{Sheykhi2011}. Furthermore, the concept of fractional gravity \cite{frac} is shown to arise if one applies the fractal Barrow entropy \cite{Barrow} to the horizon, instead of the Bekenstein-Hawking area law, suggesting therefore a profound link between these two concepts.

This alternative perspective sheds light on the possible origin of these alternative gravitational laws. It may have practical implications as well, by facilitating their implementation into the cosmological context, through their associated entropies. As demonstrated in this work, the effect of these gravitational laws on Friedmann equations can be straightforwardly addressed by appealing to the first law of thermodynamics and the specific entropic form. Another case where such a perspective may be relevant is that of fractional gravity, which is expected to involve a variable-order parameter $d(r/ \ell)$ \cite{frac}. The variable-order nature of the theory, however, makes its analytical treatment rather subtle, if possible at all. Such a program may be more easily implemented, if fractional gravity is regarded as an entropic force, by considering a Barrow-like entropy with a varying parameter $\Delta$. 

Our study opens up new avenues for future research, particularly in exploring the novel entropic forms presented here, which involve exponential or error function corrections. One possible direction is to impose observational constraints on the generalized Friedmann equations discussed in this paper (especially Eq. (\ref{FY})), checking their consistency with current cosmological data. This could involve examining the dimensionless coordinate distance data of SNe Ia and FR IIb radio galaxies, or the X-ray mass fraction data of clusters. Additionally, the novel class of entropies derived here, involving more complex corrections beyond the area law, deserves independent exploration. It would be interesting to investigate the possible constraints set on their parameters by the generalized second law of thermodynamics. Furthermore, exploring their relevance in black hole thermodynamics holds promising prospects for further understanding.

\newpage


\end{document}